\documentstyle[12pt,a4]{article}
\newcommand{\pr}{Phys.\ Rev.\ }
\newcommand{\prp}{Phys.\ Rep.\ }
\newcommand{\prep}{Phys.\ Rep.\ }

\newcommand{\np}{Nucl.\ Phys.\ }
\newcommand{\zp}{Z.\ Phys.\ }
\newcommand{\pl}{Phys.\ Lett.\ }

\newcommand{\cpc}{Comput.\ Phys.\ Commun.\ }
\def\Pom{{\bf I\!P}}
\begin{document}

\setcounter {page} {0}
\renewcommand {\thepage} { }

\renewcommand {\thefootnote} {\fnsymbol{footnote}}
\setcounter {footnote} {0}

\vspace*{1cm}
 \noindent SI 95-16

\begin{flushright}
February 1995
\end{flushright}

\vspace{5mm}

\begin{center}{\Large \bf
Hard diffraction in hadron--hadron \\ interactions and in photoproduction
}\end{center}

\vspace{5mm}
\begin{center}
R. Engel$^{*,\dag}$, J. Ranft$^{\ddag}$, S. Roesler$^{\dag}$ \\
\vspace{5mm}
{$^*$\it Universit\"at Leipzig, Fachbereich Physik, D--04109 Leipzig,
         Germany}\\
{$^{\dag}$\it Universit\"at Siegen, Fachbereich Physik, D--57068 Siegen,
         Germany}\\
{$^{\ddag}$\it CERN, CH--1211 Gen\`eve 23, Switzerland}\\
\end{center}

\vspace{3cm}
\begin{center}{\bf Abstract}\end{center}

Hard single diffractive processes are studied within the framework
of the triple--Pomeron approximation. Using a Pomeron structure
function motivated by Regge--theory we obtain parton distribution
functions which do not obey momentum sum rule. Based on Regge--
factorization cross sections for hard diffraction are calculated.
Furthermore, the model is applied to hard diffractive particle
production in photoproduction and in  $p\bar{p}$ interactions.

\vspace{2.5cm}

\vfill\newpage
\renewcommand {\thepage} {\arabic{page}}
\setcounter {page} {1}

\section {Introduction}

\noindent

A significant fraction of events in high energy hadron--hadron
interactions and in photoproduction is characterized by a
rapidity gap between a quasi--elastically scattered primary hadron and a
multiparticle final state. These single diffractive interactions can be
understood in terms of Pomeron exchange.

As suggested by Ingelman and Schlein~\cite{Ingelman85} there is now --
in addition to the well--investigated soft diffractive particle
production -- experimental evidence for
single diffractive jet production both in $p\bar{p}$ and $ep$
interactions~\cite{Bonino88,Brandt92,Ahmed94a,Ahmed94b,Derrick93a,Derrick94e}.
Characteristic
features of these jets are very similar to the ones produced in
nondiffractive events. This observation suggests that it
might be possible to
apply models based on perturbative QCD to describe hard diffractive
scattering processes between hadrons or photons and Pomerons.

It is useful to distinguish between noncoherent diffraction
and processes where the Pomeron enters the hard scattering as a whole.
Models based on
noncoherent interactions usually assume the Pomeron to be a partonic
object, the parton--momenta obeying parton distribution functions
(PDFs)~\cite
{Ingelman85,Berger87,Donnachie87,Donnachie88,Ingelman93,Nikolaev92a}.
According to the assumptions made, these PDFs strongly differ in the
partons considered, in their shapes, and in their normalizations.
In addition one can consider interactions
where almost all the Pomeron momentum enters the hard scattering. First
signs for a ``super hard'' Pomeron structure
in $p\bar{p}$ interactions
were reported by the UA8--Collaboration~\cite{Brandt92}.
These effects may manifest itself in a breakdown of factorization and
would lead to a delta--function like term in the Pomeron structure
function. They can be explained by mechanisms called ``coherent
hard diffraction''~\cite{Collins93}
or by means of a  direct Pomeron--quark
coupling~\cite{Donnachie87,Donnachie84,Kniehl94}.
However the experimental information on diffractive interactions
containing jets in the final state is still limited and does not allow
to draw definite conclusions on the Pomeron structure.

In the present paper we discuss hard diffraction in the framework of the
two--component Dual Parton Model
\cite{Capella94a,Aurenche92a,Roesler93,Engel94a}.
We use an ansatz developed by
Capella et al.~\cite{Capella94b,Capella94c}
which is based on Regge--theory to obtain the Pomeron PDFs.
Further assumptions on the normalization of the quark
distributions are not necessary since this is given by $F_2^{\Pom}$.
Features of particle production in hard diffractive photoproduction and
in hard diffractive $p\bar{p}$ interactions
are investigated and compared to data~\cite{Brandt92,Ahmed94b}.

The paper is organized as follows. In Sect.~2 we describe the way hard
diffraction is treated in our model. We especially focus on the PDFs of
the Pomeron.
Using Monte Carlo realizations of the model, we discuss hard
diffractive particle production and calculate hard diffractive cross
sections in Sect.~3.
A summary is given in Sect.~4.

\section{Description of the model}

\subsection{\label{sec_DPM}
The triple--Pomeron approximation}

Within Regge--theory, high--mass single diffractive processes are
understood by means of triple--Pomeron exchange
(see Fig.~\ref{pic_TPsoft}). Using Reggeon field
theory~\cite{Baker76} this graph is treated as an usual Feynman
graph. The Pomerons shown in Fig.~\ref{pic_TPsoft} are defined
by propagators
$\xi_{\Pom}(t)(s/s_{0})^{\alpha_{\Pom}(t)}$
and cannot be considered as particles~\cite{Kaidalov79}. In lowest
order the differential diffractive cross section is given by the
unitarity cut (Fig.~\ref{pic_TPsoft})
\begin{equation}
\label{dsigdtdm2}
\frac{d^{2}\sigma_{3\Pom}(s,t)}{dtdM^{2}}=\frac{1}{16\pi s^{2}}
|g_{11'}^{\Pom}(t)|^{2}g_{22}^{\Pom}(0)\Gamma^{3\Pom}(t,0)|\xi_{\Pom}(t)|^{2}
\left(\frac{s}{M^{2}}\right)^{2\alpha_{\Pom}(t)}
\left(\frac{M^{2}}{s_{0}}\right)^{\alpha_{\Pom}(0)}.
\label{tripom}
\end{equation}
With $M$ we denote the mass of the diffractively excited system.
$g_{11'}^{\Pom},g_{22}^{\Pom}$, and $\Gamma^{3\Pom}$ are the
various couplings as shown in Fig.~\ref{pic_TPsoft}.
$\xi_{\Pom}(t)$ is the usual signature
factor
\begin{equation}
\xi_{\Pom}(t)=-\frac{1+e^{-i\pi \alpha_{\Pom}(t)}}{\sin (\pi
\alpha_{\Pom}(t))}.
\end{equation}
$\alpha_{\Pom}(t)=1+\Delta+\alpha_{\Pom}'(0)t$ is the Pomeron
trajectory with the intercept $1+\Delta$, and $s_0=1$GeV$^2$.
Introducing an effective Pomeron--particle cross
section~\cite{Kaidalov79}
\begin{equation}
\sigma_{tot}^{a\Pom}(M^2,t)=g_{22}^{\Pom}(0)
\frac{\Gamma^{3\Pom}(t,0)}{s_0}
\left(\frac{M^2}{s_0}\right)^{\alpha_{\Pom}(0)-1},
\end{equation}
Eq.~(\ref{dsigdtdm2}) can be interpreted as a product of a Pomeron flux
factor and this cross section
\begin{equation}
\label{dsigdtdm2a}
\frac{d^{2}\sigma_{3\Pom}(s,t)}{dtdM^{2}}=\frac{1}{16\pi s^{2}}
|g_{11'}^{\Pom}(t)|^{2}|\xi_{\Pom}(t)|^{2}
\frac{s^{2\alpha_{\Pom}(t)}}{(M^{2})^{2\alpha_{\Pom}(t)-1}}
\sigma_{tot}^{a\Pom}(M^2,t).
\end{equation}
To estimate the contribution of hard diffraction (an example
of such a process is shown in Fig.~\ref{pic_TPhard}) to the single
diffractive cross section, $\sigma_{tot}^{a\Pom}$ can be replaced by
the hard Pomeron--hadron/photon cross section $\sigma_{h}^{a\Pom}$.
We obtain $\sigma_{h}^{a\Pom}$ applying lowest order perturbative
QCD, i.e. in case of hadron or resolved photon interactions
\begin{equation}
\label{sigmahres}
\sigma_h^{a\Pom}=\sum_{i,j,k,l} \frac{1}{1+\delta_{kl}} \int_0^1 dx_1
\int_0^1
dx_2 \int d\hat{t}\ \frac{d\sigma_{QCD}^{i,j\rightarrow k,l}}{d\hat{t}}\
f_{a}^i(x_1,Q^2) f_{\Pom}^j(x_2,Q^2)\ \Theta(p_\perp-p_{\perp}^{cutoff})
\end{equation}
and for direct photon--Pomeron interactions
\begin{equation}
\label{sigmahdir}
\sigma_{h,dir}^{\gamma\Pom}=\sum_{j,k,l} \int_0^1 dx
\int d\hat{t}\ \frac{d\sigma_{QCD}^{\gamma,j\rightarrow k,l}}{d\hat{t}}
f_{\Pom}^j(x,Q^2)\ \Theta(p_\perp-p_{\perp}^{cutoff}).
\end{equation}
We sum over all parton configurations and integrate having a lower
cut--off in transverse momentum $p_{\perp}^{cutoff}$.
Here $f_a^i(x,Q^2)$ are the PDFs of the hadron/photon whereas with
$f_{\Pom}^j(x,Q^2)$ we introduce distribution functions of partons inside
the Pomeron. A detailed discussion of the Pomeron--PDFs will be
given in the following section.

The free
parameters occurring in Eq.~(\ref{tripom}), such as the proton--Pomeron
coupling constant and the intercept of the Pomeron--trajectory,
are obtained within the two--component Dual Parton Model by fits to
data on total, elastic, and diffractive cross sections.
We refer to~\cite{Roesler93,Engel94a,Bopp94a,Engel92a}
for further details.

\subsection{\label{sectpdf} Parton distributions in the Pomeron}

As shown in~\cite{Donnachie87,Donnachie84,Nikolaev92a,Capella94c}
one can derive a Pomeron
structure function $F_2^{\Pom}$ by relating it to the total cross
section of virtual photon--proton  diffractive deep inelastic
scattering.
Similar to the way the proton structure function
$F_2^p$ can be related to $\sigma_{tot}^{\gamma^*p}$ one gets
\begin{equation}
\label{deff2po}
F_2^{\Pom}(x,Q^2,t) = \frac{Q^2}{4\pi^2\alpha_{em}}
\sigma_{tot}^{\gamma^*\Pom}(M^2,Q^2,t),\qquad x=\frac{Q^2}{Q^2+M^2}.
\end{equation}
Convoluting
$F_2^{\Pom}$ with a Pomeron--flux factor, one obtains the structure function
of diffractive dissociation. Using Regge
factorization it is possible to calculate $F_2^{\Pom}$ from the deuteron
structure function $F_2^d$. The complete formalism is given
in Ref.~\cite{Capella94b,Capella94c}.
Here we only want to give the main formulas
which are necessary to understand the PDFs of the Pomeron we will use
afterwards. The proton structure function is parametrized at moderate
values of $Q^2$ ($Q^2\le 5$~GeV$^2$)~\cite{Capella94b}
\begin{eqnarray}
\label{f2prot}
\lefteqn{F_2^p(x,Q^2)=S(x,Q^2)+V(x,Q^2)=} \nonumber\\
& &A x^{-\Delta(Q^2)} (1-x)^{n(Q^2)+4} \left(
\frac{Q^2}{Q^2+a}\right)^{1+\Delta(Q^2)}  \nonumber\\
& &+B\ x^{1-\alpha_R} (1-x)^{n(Q^2)} \left(
\frac{Q^2}{Q^2+b}\right)^{\alpha_R}
\end{eqnarray}
with
\begin{equation}
\Delta(Q^2) = \Delta_0\left(1+\frac{2Q^2}{Q^2+d}\right), \qquad
n(Q^2) = \frac{3}{2} \left(1+\frac{Q^2}{Q^2+c}\right).
\end{equation}
We again refer to~\cite{Capella94b,Capella94c} for the exact values of
the parameters entering the expressions and a detailed discussion of the
$Q^2$--dependent intercept.
The Pomeron structure function can be related to the deuteron structure
function by the following substitutions
\begin{equation}
\label{f2po}
F_2^{\Pom}(x,Q^2,t) = F_2^d(x,Q^2;A\rightarrow e(t)A,B\rightarrow f(t)B,
n(Q^2)\rightarrow n(Q^2)-2).
\label{subst}
\end{equation}
$e$ and $f$ are ratios of coupling constants.
The $t$--dependence of the Pomeron--PDF is completely given by
the ratios $e$ and $f$.
We use $e(0)=3f(0)=0.1$~\cite{Capella94c}
and an exponential dependence on $t$ with the slope
$b=0.5$~GeV$^{-2}$~\cite{Kaidalov79,Engel94a,Capella75}.
The substitution $n(Q^2)\rightarrow n(Q^2)-2$, for example, is due
to the similarity of the valence quark distribution in the Pomeron
and a meson.
The first
term in Eq.~(\ref{f2prot}) determines the sea-quark distribution whereas
the last term is responsible for the behaviour of the valence
quark distribution. Using Eq.~(\ref{f2po}), and the definition
\begin{equation}
F_2^{\Pom}(x,Q^2,t) = \sum_q e_q^2 x (f_{\Pom}^q(x,Q^2;t)+
f_{\Pom}^{\bar{q}}(x,Q^2;t))
\end{equation}
where $e_q$ are the corresponding quark--charges, we obtain at
moderate $Q^2$--values for the Pomeron PDFs
\begin{eqnarray}
\label{popdfq}
xf_{\Pom}^u(x,Q^2;t)&=&xf_{\Pom}^{\bar{u}}(x,Q^2;t) = xf_{\Pom}^d(x,Q^2;t) =
xf_{\Pom}^{\bar{d}}(x,Q^2;t) \nonumber \\
&=& \frac{3}{4} e(t)S(x,Q^2)+\frac{9}{10} f(t)V(x,Q^2)
\nonumber \\
xf_{\Pom}^s(x,Q^2;t)&=&xf_{\Pom}^{\bar{s}}(x,Q^2;t) =
\frac{3}{4}e(t)S(x,Q^2).
\end{eqnarray}
For simplicity, we assume a
SU(3)--symmetrical sea of light quarks. The charm--quark
distribution of the Pomeron is suppressed at the $Q^2$--values
considered so far.
In contrast to the nucleon it is not possible to determine the
normalization of the gluon distribution in the Pomeron by momentum sum rule.
We are therefore free to choose a specific form and normalization.
With respect to the experimental observations which
favor a relatively ``hard'' structure we use
\begin{equation}
\label{popdfg}
xf_{\Pom}^g(x,Q_0^2;t) = K(Q_0^2,t) x(1-x)
\end{equation}
where $K(Q_0^2,t)$ depends on the normalization chosen. In the following we
take the normalization of the gluon distribution according to the
scaling factor resulting for the quark distribution from Eq.~(\ref{subst}).

As already mentioned all distributions obtained are limited to $Q^2\le
5$~GeV$^2$. In order to get PDFs of the Pomeron at higher $Q^2$--values
we take (\ref{popdfq}) and (\ref{popdfg}) as input distributions for a
QCD evolution in the leading logarithmic approximation~\cite{Nikolaev92a}.
The evolution was done using the code of~\cite{Devoto83,Aurenche89} with
$Q_0^2=2$~GeV$^2$. The result is shown in
Fig.~\ref{pic_popdf1} and \ref{pic_popdf2}, where we have plotted
$xf_{\Pom}^u, xf_{\Pom}^s$, and $xf_{\Pom}^g$ for different values of $Q^2$.
Within this formalism, the normalization of the Pomeron PDFs is given by
\begin{equation}
N_q=\sum_{q,\bar{q}}\int_0^1 dx\ xf_{\Pom}^q(x,Q^2,t),\qquad
N_g=\int_0^1 dx\ xf_{\Pom}^g(x,Q^2,t)
\end{equation}
which is shown for $t=0$ GeV$^2$ in Fig.~\ref{pic_norm}.
The hardness of the $u$-- and $d$--quark distributions at low values of
$Q^2$ is governed by the valence part of~(\ref{f2po}), i.e. for $x>0.5$
we are mainly dealing with a $\sqrt{x}(1-x)^{n(Q^2)-2}$--behaviour,
$n(Q^2)$ being
about 2.0...3.0 ($c=3.55$~\cite{Capella94b}). We note that this behaviour
is similar to the $Q^2$--independent predictions
of~\cite{Donnachie87,Donnachie84}, whereas at high $Q^2$--values our
quark--distributions become softened due to the sea-quark contribution
and the QCD--evolution.

Furthermore, we show in Fig.~\ref{pic_f2diff} the structure function
$F_2^{\Pom}$ for different $Q^2$--scales. Again, the flat shape
of $F_2^{\Pom}$ is determined by the valence quark distributions.

\section{Particle production in hard diffractive interactions}

\subsection{Sampling of hard single diffractive events}

The Monte--Carlo treatment of soft single diffractive hadron--hadron
interactions within the two--component DPM is described
in~\cite{Aurenche92a,Roesler93}.
The generation of single diffractive events in
photoproduction will be discussed elsewhere~\cite{Engel95ip}. Here we
want to focus on the way the existing models have been extended to
diffractive jet--production.
This has been done for $p\bar{p}$ collisions using an extension
to the DTUJET-93 code~\cite{Bopp94b} and for $p\bar{p}$ as well as
$\gamma p$ collisions using the DTUJET-PHOJET code~\cite{Engel94c,Engel95ip}.

The Monte--Carlo implementation is similar to that of usual hard
scattering processes (i.e. processes involving high $p_t$) between
hadrons or between hadrons and photons.
The main differences are: (i) the
interaction is boosted to the rest system of the diffractively excited
``blob'', the  CM  energy is therefore given by the diffractive
mass, and (ii) one hadron is replaced by a Pomeron with a
virtuality $t$.
The momenta of the partons
entering the hard $2\longrightarrow 2$ scattering process are obtained
using conventional PDFs of hadrons and photons~\cite{MRS93,GRV92b,PDFLIB93}
and the
PDFs of the Pomeron introduced in the previous section. Initial state
radiation which significantly modifies the multiparticle final state of
interactions involving high transverse momenta has been implemented. The
chain system to be hadronized using
JETSET~7.3~\cite{Sjostrand86,Sjostrand87a} is determined by the color
flow taking cross sections for different color flow diagrams into
consideration~\cite{Combridge77,Bengtsson84}.

Parts of the MC--realization of hard diffraction are technically similar
to the ones described in~\cite{Bruni93,Jung93a},
which were so far mainly used to
understand the underlying interactions. However, we would like to
emphasize that our starting points are completely different. Our
investigation is based on the two--component DPM which treats soft and
hard scattering processes in an unified manner. The free parameters are
fixed by fits to cross section data. Similar to~\cite{Bruni93,Jung93a}
we assume hard
diffraction to be based on a partonic structure of the Pomeron but we
start from a Pomeron structure function which is completely determined
by fits to data on the proton structure function and a ratio of coupling
constants which follows from the model. There is no further freedom in
choosing a specific $x$-- and $Q^2$--dependence of the quark
distribution inside the Pomeron. The normalizations of the Pomeron PDFs
(and therefore the hard diffractive cross sections) are obtained from
the model rather than imposing additional
assumptions~\cite{Bruni93,Jung93a}.

\subsection{Hard diffraction in photoproduction}

Recent measurements at the electron proton collider HERA at DESY
have shown that
a substantial part of minimum bias photoproduction \cite{Ahmed94b,Derrick94e}
and deep inelastic
scattering \cite{Ahmed94a,Derrick93a} events exhibits diffractive
features similar to hadron--hadron
scattering.
First distributions of so called rapidity gap events have been
published
\cite{Ahmed94b,Derrick94e},
but the data are not yet corrected for acceptance and do not allow
absolute comparisons.
We will apply the model developed
to investigate diffraction dissociation  of photons
in quasi--real photon--hadron scattering and we understand that
more data will be published soon.

The simulation of diffractive events in photoproduction can be done
similar to diffraction in hadron--hadron scattering substituting
the hadron--Pomeron scattering subprocess by photon--Pomeron scattering.
In the calculation of the total diffractive cross section, the absorptive
corrections due to multiple photon-hadron scattering
are taken into account \cite{Engel94a}. For the simulation of
photon-Pomeron scattering  we neglect unitarity corrections
(e.g.\ multiple photon-Pomeron scattering)
which become important only for very high diffractive masses\footnote{
High diffractive masses are suppressed by applying the experimental cuts.}.
Thus, the simulation of hard photon--Pomeron scattering follows directly from
Eq.~(\ref{sigmahres},\ref{sigmahdir}).
The Pomeron flux is calculated using Eq.~(\ref{tripom}).

To calculate $ep$ photoproduction cross sections, the flux of
quasi-real photons has to be estimated.
Here we apply the
improved Weizs\"acker--Williams approximation
\cite{Frixione93}
\begin{equation}
f_{e/\gamma}(y) = \frac{\alpha}{2 \pi}
\left( \frac{1+(1+y)^2}{y} \mbox{ln} \frac{Q^2_{max}}{Q^2_{min}}
- \frac{1 (1-y)}{y} \right)
\end{equation}
with the kinematical cuts
\begin{equation}
0.25 < y < 0.7 \hspace*{1cm} Q^2_{min} = 3\cdot 10^{-8} \mbox{GeV}^2
\hspace*{1cm} Q^2_{max} = 10^{-2} \mbox{GeV}^2.
\end{equation}
$y$ and $Q^2$ denote
the energy fraction taken by the photon from the electron
and the photon virtuality, respectively.

Using our Monte Carlo program, complete hadronic final states have been
generated and analyzed.
According to the experimental conditions only events passing the
$\eta_{max}$-cut \cite{Ahmed93a,Derrick93a}
have been accepted for the further analysis.
In addition, a few less
restrictive cuts have been applied to match the experimental selection
procedure described in \cite{Ahmed94b}.

In Fig.~\ref{pic_diffpt} we show the transverse momentum distribution
of charged particles for the selected rapidity gap events in the
pseudorapidity range $-1.5 < \eta_{lab} < 1.5$.
The absolute cross sections
obtained with the model are given. Together with the predictions we
show uncorrected data of the H1 Collaboration \cite{Ahmed94b} on
charged tracks. It is expected that the shape of this distribution
can be compared with calculations for transverse momenta  higher than
1 GeV/c \cite{Rostovtsev94pc}.
Note that the data are scaled in order to compare their shape to our
calculation.
The systematic difference between the
model and the calculation at low $p_\perp$ can be
qualitatively explained by the $p_\perp$-dependent experimental
acceptance in this region~\cite{Rostovtsev94pc}.
In addition, the contribution from direct photon processes is
shown.

Using a cone jet algorithm similar to the one used in~\cite{Ahmed94b}
the transverse energy and pseudorapidity distributions
of jets with $E_t > 4$ GeV are calculated. In
Fig.~\ref{pic_diffet} we show the transverse energy distribution of
jets together with uncorrected H1 data \cite{Ahmed94b}.
Since the so called jet-pedestal is small in diffractive
events~\cite{Ahmed94b,Derrick94e,Schlein94talk}
the transverse jet distribution
should not be drastically influenced by acceptance effects.
In Fig.~\ref{pic_diffpr} the corresponding pseudorapidity distribution
of jets is compared to H1 data. In both figures, the H1 data are again
scaled to compare their shape with the model predictions.

\subsection{Hard diffractive proton--antiproton interactions}

\subsubsection{Cross sections for hard single diffraction}

An estimation of cross sections for hard diffractive events containing
jets in the final state can be obtained using Regge--factorization
(Eq.~(\ref{dsigdtdm2a})) together with the hard
Pomeron--particle cross sections given in Eq.~(\ref{sigmahres})
and (\ref{sigmahdir}). However the cross sections depend
strongly on the partonic cut--off in transverse momentum which enters the
integrations. In addition, further uncertainties arise from choosing a
certain scale in (\ref{sigmahres}) and (\ref{sigmahdir}) and from
the definition of a diffractive event itself, i.e. from the $t$-- and
$M^2$--ranges the differential hard diffractive cross section
has to be integrated over.

Reliable predictions for cross sections in hard diffraction can
therefore only be given for a certain experimental set--up taking into
account all kinematical cuts applied and jet rates based on jet--finding
algorithms which were used to obtain the experimental results.
This has been done
for the experimental set--up of UA8~\cite{Brandt92} with a jet--finding
algorithm which will be described further below.
We calculate the ratio of the hard diffractive cross section
and the total diffractive cross section using the
MRS~D0$^\prime$ \cite{MRS93} set for the parton distributions in the
proton/antiproton and
$g_{p\Pom}(0)=6.2$~$\sqrt{mb}$,
$\Gamma^{3\Pom}(t=0)=0.08$~$\sqrt{mb}$~GeV$^2$,
$\Delta = 0.078$, $\alpha'(0)=0.25$~GeV$^{-2}$ and
$Q^2=p_{\perp}^2$~\cite{Engel94a}.
To obtain ratios which correspond to the UA8--cuts
we multiply these values by
the fraction of those events which contain at least two jets of a
transverse energy
$E_t^{jet}\ge E_t^{cutoff}$ and of a pseudorapidity
$|\eta|<2$ in their final state.
Experimental data for these ratios in preliminary form were given in
\cite{Schlein93}, final values will be available soon \cite{Brandt95a}
and we understand, that our calculations shown
in Fig.~\ref{pic_xsratio} will be consistent with these
data~\cite{Schlein94pc}.

\subsubsection{Hard diffractive particle production}

Hard diffractive proton--antiproton interactions were recently
investigated by UA8--Collaboration~\cite{Brandt92} at a  CM  energy of
630 GeV.
Only limited information on absolute hard diffractive cross sections is
available so far~\cite{Schlein93} (see previous section).
It is therefore not possible to compare our model directly to these
data. However, since our model is able to describe data on hard
diffraction in photoproduction rather well it might be worthwhile to
show also predictions for hard diffractive $p\bar{p}$--interactions.
In particular we apply the same cuts to the final
state as they were used in the experiments~\cite{Brandt92}.
Jets are identified using a cone--algorithm in the
$\eta-\phi$--plane, with $\phi$ being the azimuthal angle and
$|\eta|<2$. If a jet with
$E_t>8$ GeV is found the search for its axis is iterated
calculating $E_t$--weighted sums over cells within an unit--cone
radius. According to the UA8--data all distributions discussed in this
section are normalized to unit area.
Our results are obtained with the PDF set MRS~D0' for the parton
distributions in the proton/antiproton.

As mentioned in~\cite{Brandt92} a variable sensitive to the partonic
structure of the Pomeron could be $x_{jet}$. It is defined as the
longitudinal momentum component of a jet normalized to its maximum value
in the Pomeron--antiproton  CM  system. In Fig.~\ref{pic_x1jet} we
present predictions of our model for momentum fractions
of the quasi--elastically scattered proton $x_p$ between 0.92 and
0.94. As stated by UA8~\cite{Brandt92} the
data were obtained with essentially full acceptance at positive
$x_{jet}$--values, i.e. it can be expected that our calculations
agree well with the data also after the corrections have been applied.

In Fig.~\ref{pic_eta1jet} we show the distribution of the
jet--pseudorapidity in the antiproton--proton CM system again for
the same $x_p$--bin. The tail at positive $\eta$--values is influenced
mainly by the structure of the Pomeron.

A variable which may indicate whether there is a ``super hard'' Pomeron
structure is $x_{2jets}$ -- the longitudinal momentum of a two--jet
system, again normalized to its maximum value in the Pomeron--antiproton
system. The results of our calculation are plotted in
Fig.~\ref{pic_x2jet} showing the contribution from quarks and gluons
of the Pomeron--PDF separately.
Due to the shape of the valence quark distribution
(see Sect.~\ref{sectpdf}) the quarks mainly contribute to higher
values of $x_{2jets}$ whereas the gluon contribution is peaked around
0.2. Using our Pomeron-PDF we therefore obtain a significant fraction
of events containing 2--jet systems with $x_{2jets}>0.7$.
Considering the reasonable agreement found with the $\gamma p$ data
the question to which extent a ``super hard'' contribution is still
necessary within our model cannot be answered before corrected
measurements are available.

\section{Summary and conclusions}

Hard diffraction in hadron--hadron collisions and in photoproduction has
been investigated in the framework of the two--component DPM. Since soft
diffractive interactions at collider energies are well--described in
terms of Pomeron exchange processes, diffractive jet--production may
provide new information on the Pomeron--structure~\cite{Ingelman85}.

In the present paper
we study features of hard diffractive particle production treating the
Pomeron as a partonic object.
PDFs of the Pomeron\footnote{The code is available on request from
the authors (email: sroesler@cernvm.cern.ch).}
are obtained from a Pomeron structure function motivated by
Regge--theory. Whereas the quark distributions follow directly from the
parametrization of the Pomeron structure function, it is possible to
choose an ansatz for the gluon distribution which accounts for the
experimentally observed hard Pomeron structure. The normalizations of
the quark distributions of the Pomeron are determined by the
scaling factor relating the Pomeron structure function to the deuteron
structure function. The Pomeron PDFs are evolved to high values of $Q^2$
applying leading logarithmic QCD evolution equations.
Using the triple-Pomeron approximation, we are able to give
predictions on absolute cross sections and distributions.

We  demonstrate that our model and the Pomeron PDFs used
are in reasonable agreement with presently available data on hard
diffraction in $\gamma p$ collisions from the HERA--collider. Since there
are no absolute distributions published the comparison was restricted
to the shape.

Cross sections for hard diffraction depend strongly on kinematical cuts
and on assumptions defining an event as being produced diffractively.
Predictions can therefore only be given taking a specific experimental
set--up into account.

A comparison of hard diffractive particle production in hadron--hadron
interactions to data is presently limited to data published by
the UA8--Collaboration which do
not allow absolute comparisons. Nevertheless, the results
obtained within our model look promising and may explain the main
features of these processes.
We are not yet able to draw any conclusions concerning a possible
``super hard'' Pomeron structure.
Further investigations will be necessary as soon as more
data become available.

\section*{Acknowledgements}

The authors acknowledge stimulating discussions with
F.W.\ Bopp, A.\ Capella, C.\ Merino, D.\ Pertermann, and P.\ Schlein.
We would like to thank P. Aurenche providing us the QCD--evolution code.
We are grateful to G.\ Ingelman  and P.\ Schlein for helpful discussions
concerning the UA8 data. One of the authors (R.E.)
is indebted to  A.\ Rostovstev (H1 Collaboration) for many discussions
and valuable information on the HERA data.


\clearpage

\clearpage
\section*{Figure Captions}
\begin{enumerate}
\item \label{pic_TPsoft} A cut triple--Pomeron graph (a) describes
      multiparticle final states characterized by a rapidity gap (b).
\item \label{pic_TPhard} Example for a hard diffractive scattering
      process. A gluon of the Pomeron undergoes a hard
      scattering with a gluon of the lower particle labeled with 2.
\item \label{pic_popdf1} $u$-- and $s$--quark distributions in the
      Pomeron are plotted for different values of $Q^2$. The initial
      distribution of the QCD--evolution is shown for
      $Q_0^2=2$~GeV$^2$.
\item \label{pic_popdf2} Gluon distribution in the
      Pomeron shown for different values of $Q^2$.
\item \label{pic_norm} Normalization of the quark and gluon distribution
      function of the Pomeron.
\item \label{pic_f2diff} Pomeron structure function
      $\protect{F_2^\Pom(x,Q^2;t=0)}$.
\item \label{pic_diffpt} Inclusive charged particle cross
      section for particle with $|\eta_{lab}| < 1.5$.
      The model predictions are
      shown as full line and compared to H1 data (see text).
\item \label{pic_diffet} Inclusive transverse energy
      distribution of jets calculated for jets with $|\eta_{jet}| < 1.5$
      and compared to H1 data.
\item \label{pic_diffpr} Inclusive pseudorapidity distribution
      of jets in rapidity gap events with $E_{t,jet} > 4$ GeV
      calculated with the model and shown with H1 data.
\item \label{pic_xsratio}
      Ratio of hard diffractive cross sections for two--jet events and
      total diffractive cross sections.
      According to forthcoming the UA8
      data~\protect\cite{Schlein94pc,Brandt95a}
      to whom our calculation might be compared,
      a lower cut in transverse energy of 8 GeV has been applied.
\item \label{pic_x1jet} Distribution of the longitudinal momenta of
      jets normalized to their maximum value
      in the  CM  system of the Pomeron--proton interaction.
      The predictions of the model (labeled ``MC'')
      are shown for a momentum fraction
      of the quasi-elastically scattered proton between 0.92 and
      0.94. The uncorrected UA8--data~\protect\cite{Brandt92}
      are plotted separately (labeled ``UA8'').
\item \label{pic_eta1jet}
      Pseudorapidity distributions of jets in the CM system of the
      proton--antiproton interaction. The Monte Carlo results (MC)
      are shown for a momentum fraction
      of the quasi-elastically scattered proton between 0.92 and
      0.94. The uncorrected UA8--data~\protect\cite{Brandt92}
      are plotted separately.
\item \label{pic_x2jet} Distributions of the total longitudinal momentum
      of a two--jet system normalized to the maximum value
      for the quark and gluon contribution of the Pomeron
      and both contributions together
      are shown. Separately we plot the uncorrected
      UA8--data~\protect\cite{Brandt92}.

\end{enumerate}


\clearpage
\begin{thebibliography}{10}

\bibitem{Ingelman85}
G.~Ingelman and P.~E. Schlein:
\newblock \pl B152 (1985) 256

\bibitem{Bonino88}
UA8 Collab.:  R.~Bonino et~al.:
\newblock \pl B211 (1988) 239

\bibitem{Brandt92}
UA8 Collab.:  A.~Brandt et~al.:
\newblock \pl B297 (1992) 417

\bibitem{Ahmed94a}
H1 Collab.:  T.~Ahmed et~al.:
\newblock \np B429 (1994) 477

\bibitem{Ahmed94b}
H1 Collab.:  T.~Ahmed et~al.:
\newblock Observation of hard processes in rapidity gap events in $\gamma p$
  interactions at HERA,
\newblock DESY 94-198,
\newblock to be published in \np,  1994

\bibitem{Derrick93a}
ZEUS Collab.:  M.~Derrick et~al.:
\newblock \pl B315 (1993) 481

\bibitem{Derrick94e}
ZEUS Collab.:  M.~Derrick et~al.:
\newblock Observation of hard scattering in photoproduction events with a
  large rapidity gap at HERA,
\newblock DESY 94-210,
\newblock 1994

\bibitem{Berger87}
E.~L. Berger, J.~C. Collins, D.~E. Soper  and G.~Sterman:
\newblock \np B286 (1987) 704

\bibitem{Donnachie87}
A.~Donnachie and P.~V. Landshoff:
\newblock \pl B191 (1987) 309

\bibitem{Donnachie88}
A.~Donnachie and P.~V. Landshoff:
\newblock \np B303 (1988) 634

\bibitem{Ingelman93}
G.~Ingelman and K.~Prytz:
\newblock \zp C58 (1993) 285

\bibitem{Nikolaev92a}
N.~N. Nikolaev and B.~G. Zakharov:
\newblock \zp C53 (1992) 331

\bibitem{Collins93}
J.~Collins, L.~Frankfurt  and M.~Strikman:
\newblock \pl B307 (1993) 161

\bibitem{Donnachie84}
A.~Donnachie and P.~V. Landshoff:
\newblock \np B244 (1984) 322

\bibitem{Kniehl94}
B.~Kniehl, H.-G. Kohrs  and G.~Kramers:
\newblock Diffractive photoproduction of jets with a direct Pomeron
  coupling at HERA,
\newblock DESY 94-140,
\newblock 1994

\bibitem{Capella94a}
A.~Capella, U.~Sukhatme, C.~I. Tan  and J.~Tran Thanh~Van:
\newblock \prp 236 (1994) 227

\bibitem{Aurenche92a}
P.~Aurenche, F.~W. Bopp, A.~Capella, J.~Kwiecinski, M.~Maire, J.~Ranft  and
  J.~Tran Thanh~Van:
\newblock \pr D45 (1992) 92

\bibitem{Roesler93}
S.~Roesler, R.~Engel  and J.~Ranft:
\newblock \zp C59 (1993) 481

\bibitem{Engel94a}
R.~Engel:
\newblock Photoproduction within the two-component Dual Parton Model:
  Amplitudes and Cross Sections,
\newblock Siegen University SI 94-02,
\newblock to be published in \zp C,  1994

\bibitem{Capella94b}
A.~Capella, A.~Kaidalov, C.~Merino  and J.~Tran Thanh~Van:
\newblock \pl B337 (1994) 358

\bibitem{Capella94c}
A.~Capella, A.~Kaidalov, C.~Merino  and J.~Tran Thanh~Van:
\newblock Diffractive dissociation in deep inelastic scattering at HERA,
\newblock LPTHE Orsay 94-42,
\newblock to be published in \pl\ B,  1994

\bibitem{Baker76}
M.~Baker and K.~A. Ter-Martirosyan:
\newblock \prp 28C (1976) 1

\bibitem{Kaidalov79}
A.~B. Kaidalov:
\newblock \prep 50 (1979) 157

\bibitem{Bopp94a}
F.~W. Bopp, R.~Engel, D.~Pertermann  and Ranft:
\newblock \pr D49 (1994) 3236

\bibitem{Engel92a}
R.~Engel, F.~W. Bopp, D.~Pertermann  and J.~Ranft:
\newblock \pr D46 (1992) 5192

\bibitem{Capella75}
A.~Capella, J.~Tran Thanh~Van  and J.~Kaplan:
\newblock \np B97 (1975) 493

\bibitem{Devoto83}
A.~Devoto, D.~W. Duke  and J.~F. Owens:
\newblock \pr D27 (1983) 508

\bibitem{Aurenche89}
P.~Aurenche, R.~Baier, M.~Fontannaz, M.~N. Kienzle-Focacci  and M.~Werlen:
\newblock \pl B233 (1989) 517

\bibitem{Engel95ip}
R.~Engel:
\newblock {in preparation},

\bibitem{Bopp94b}
F.~W. Bopp, R.~Engel, D.~Pertermann, J.~Ranft  and S.~Roesler:
\newblock \cpc 83 (1994) 107

\bibitem{Engel94c}
R.~Engel:
\newblock Photoproduction within the Dual Parton Model,
\newblock {to be published in Proceedings of Rencontre de Moriond, Meribel},
  1994

\bibitem{MRS93}
A.~D. Martin, R.~G. Roberts  and W.~J. Stirling:
\newblock \pl B306 (1993) 145

\bibitem{GRV92b}
M.~Gl\"uck, E.~Reya  and A.~Vogt:
\newblock \pr D46 (1992) 1973

\bibitem{PDFLIB93}
H.~Plothow-Besch:
\newblock \cpc 75 (1993) 396

\bibitem{Sjostrand86}
T.~Sj\"ostrand:
\newblock \cpc 39 (1986) 347

\bibitem{Sjostrand87a}
T.~Sj\"ostrand and M.~Bengtsson:
\newblock \cpc 43 (1987) 367

\bibitem{Combridge77}
B.~L. Combridge, J.~Kripfganz  and J.~Ranft:
\newblock \pl B70 (1977) 234

\bibitem{Bengtsson84}
H.~U. Bengtsson:
\newblock \cpc 31 (1984) 323

\bibitem{Bruni93}
P.~Bruni and G.~Ingelman:
\newblock Diffractive hard scattering at $ep$ and $pp$ colliders,
\newblock Talk given at Marseille HEP Conference, in Proccedings of
  International European Conference on High Energy Physics, Page 595, ed.\ by
  J.~Carr-M.~Perrottet, Edition Frontieres, Gif-sur-Yvette,  1994

\bibitem{Jung93a}
H.~Jung:
\newblock Hard diffractive scattering in high energy $ep$ collisions and the
  Monte Carlo generator RAPGAP,
\newblock DESY 93-182,
\newblock submitted to \cpc,  1993

\bibitem{Frixione93}
S.~Frixione, M.~Mangano, P.~Nanson  and G.~Ridolfi:
\newblock \pl B319 (1993) 339

\bibitem{Ahmed93a}
H1 Collab.:  T.~Ahmed et~al.:
\newblock \pl B299 (1993) 374

\bibitem{Rostovtsev94pc}
A.~Rostovtsev:
\newblock private communication,

\bibitem{Schlein94talk}
P.~E. Schlein:
\newblock {Talk given at the 14th Meeting of the Large Hadron Collider
  Committee, CERN},
\newblock 1994

\bibitem{Schlein93}
P.~E. Schlein:
\newblock The evidence for partonic behaviour of the Pomeron,
\newblock preprint UCLA-PPh0061,
\newblock Talk given at Marseille HEP Conference, in Proccedings of
  International European Conference on High Energy Physics, Page 592, ed.\ by
  J.~Carr-M.~Perrottet, Edition Frontieres, Gif-sur-Yvette,  1994

\bibitem{Brandt95a}
UA8 Collab.:  A.~Brandt et~al.:
\newblock Cross section measurement of hard diffraction at the SPS-Collider,
\newblock in preparation,  1995

\bibitem{Schlein94pc}
P.~E. Schlein:
\newblock private communication,

\end{thebibliography}
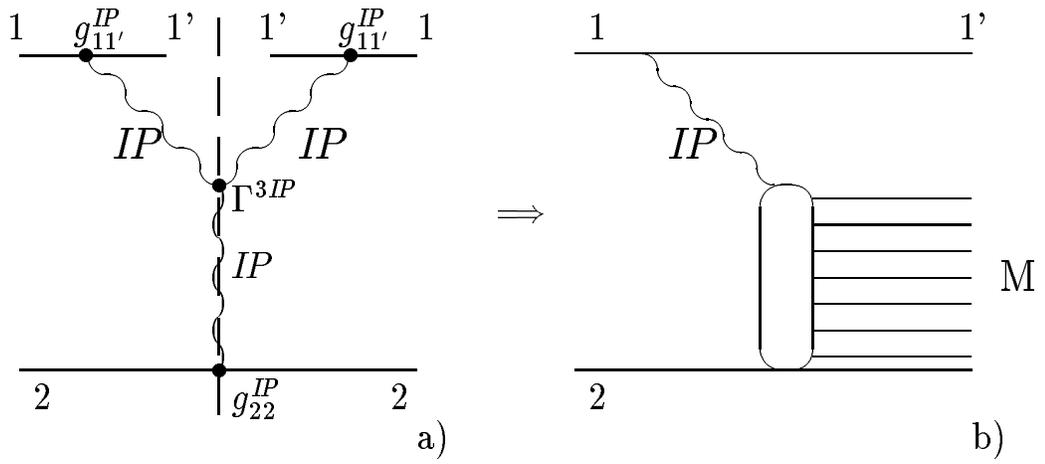
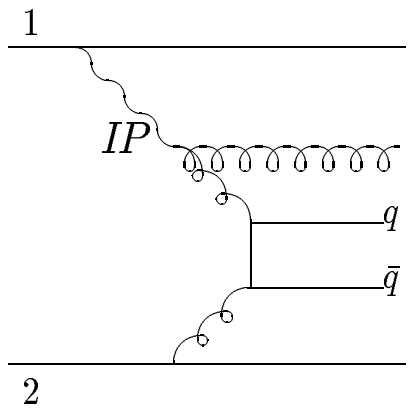
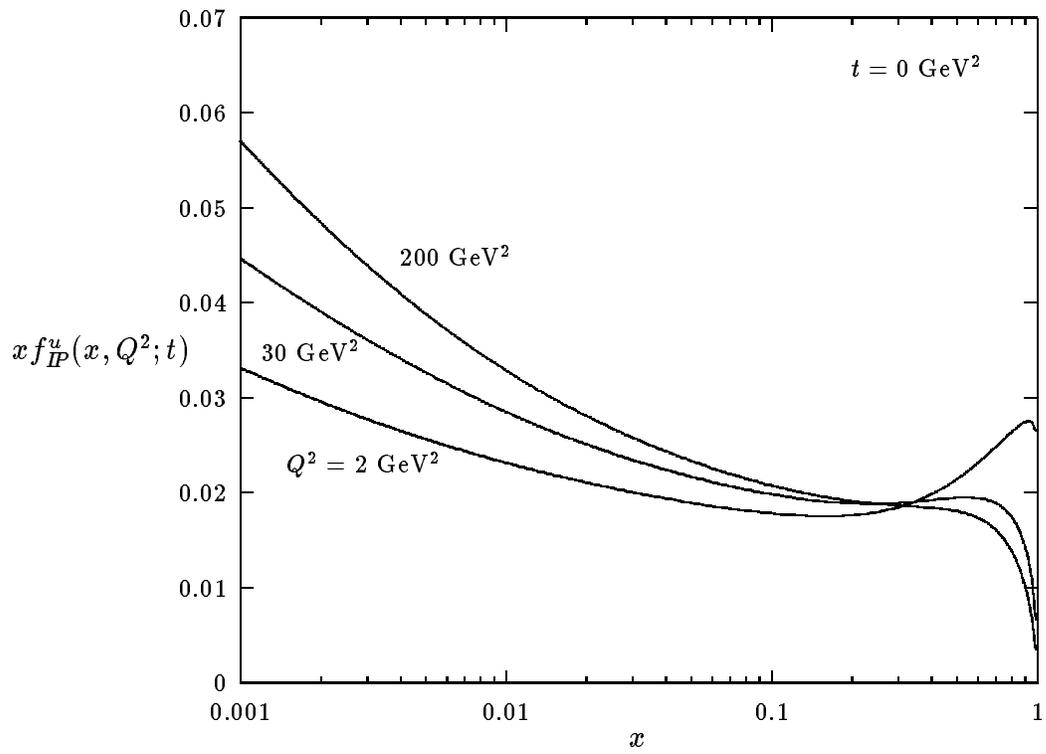
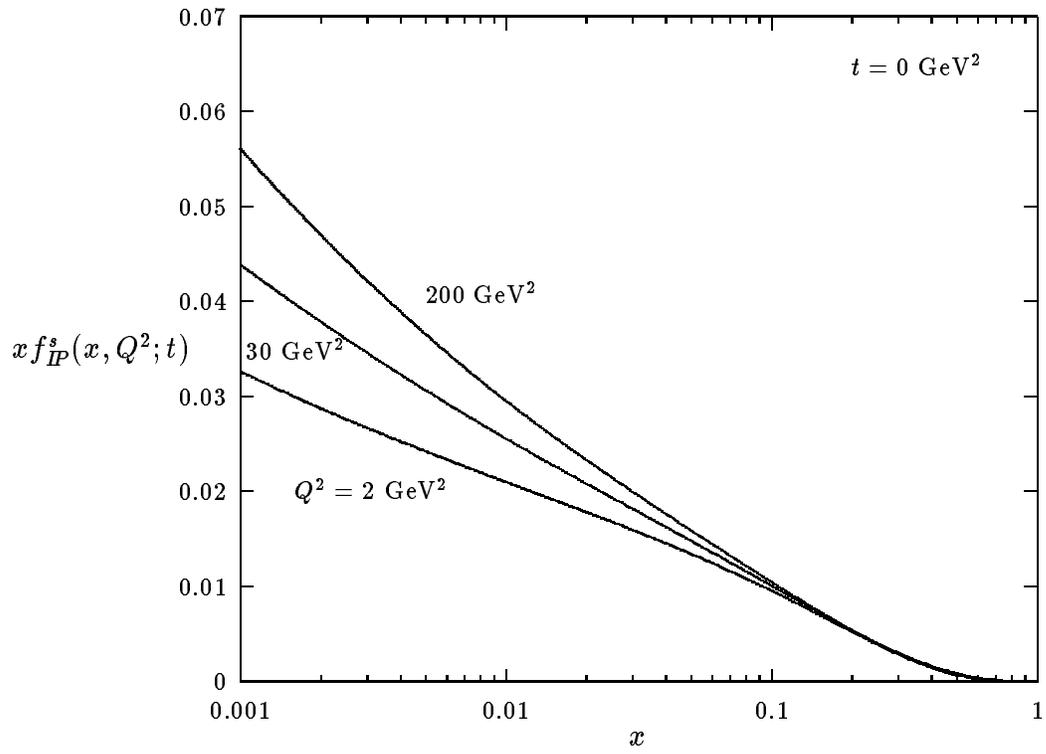
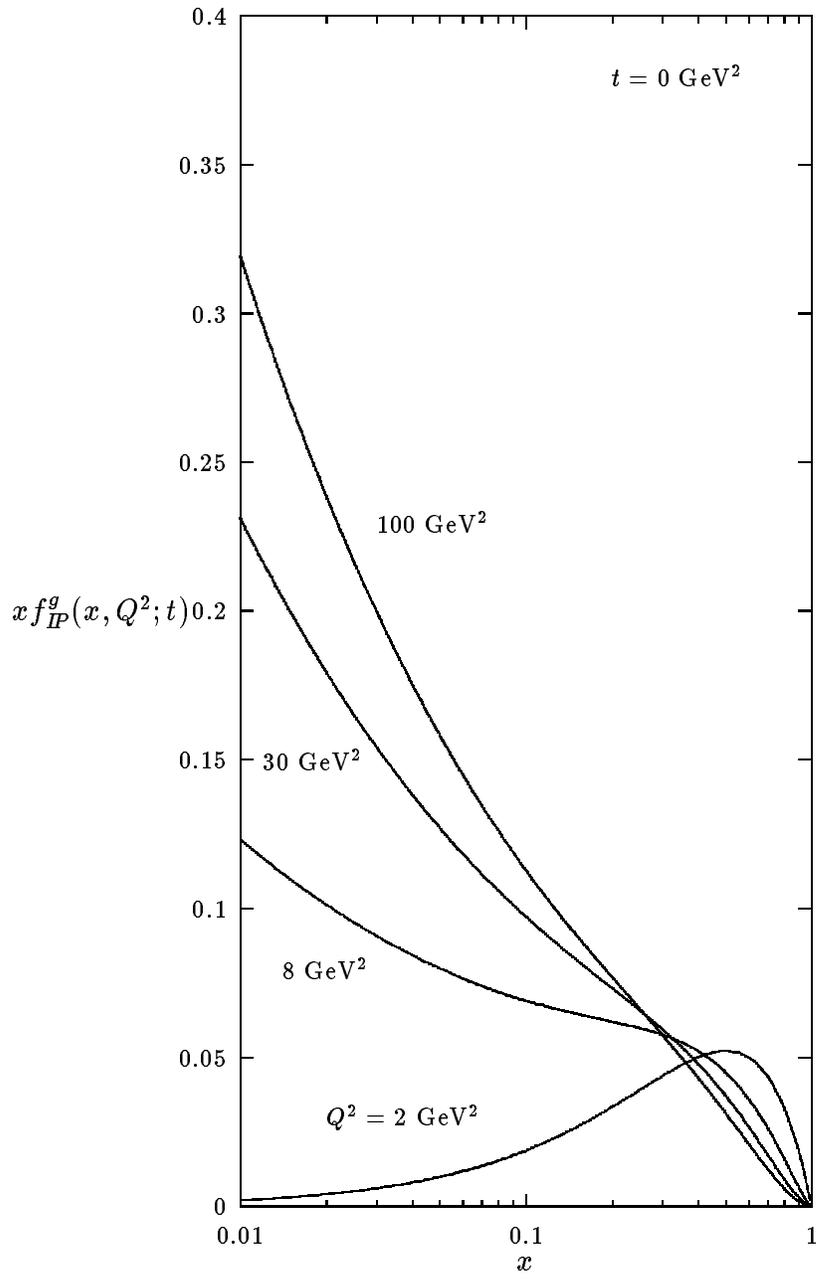
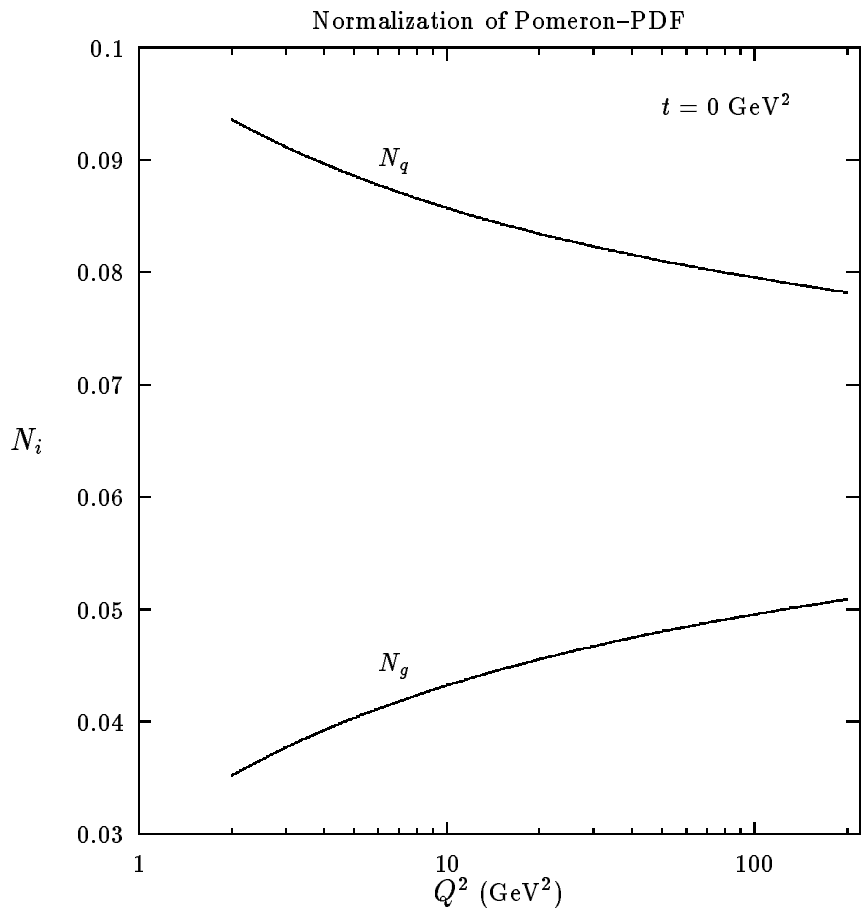
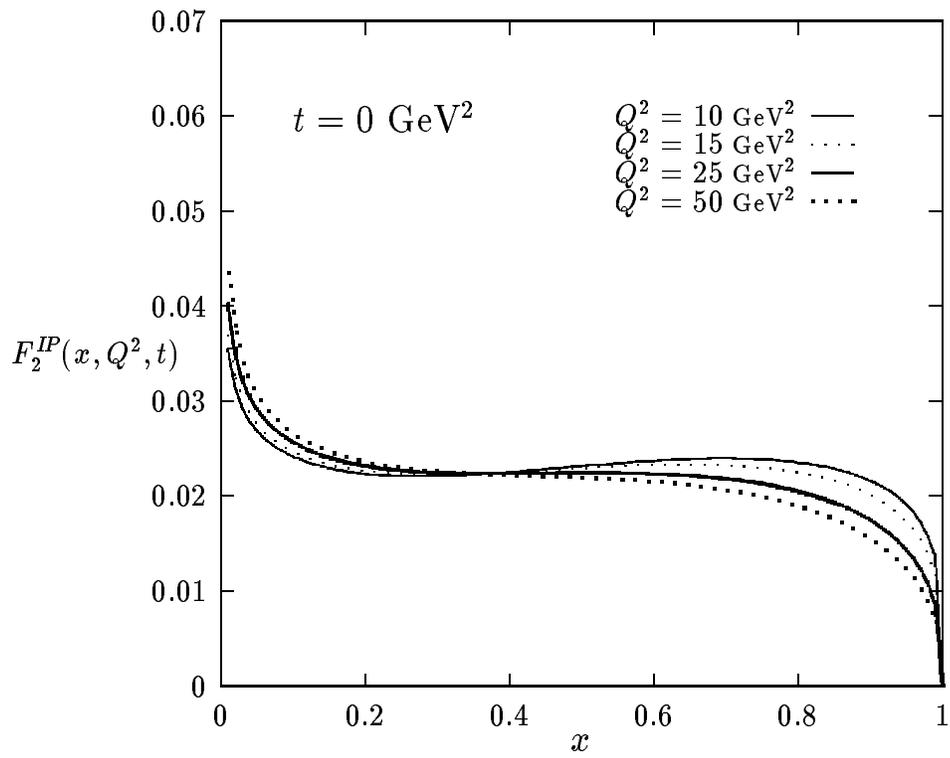
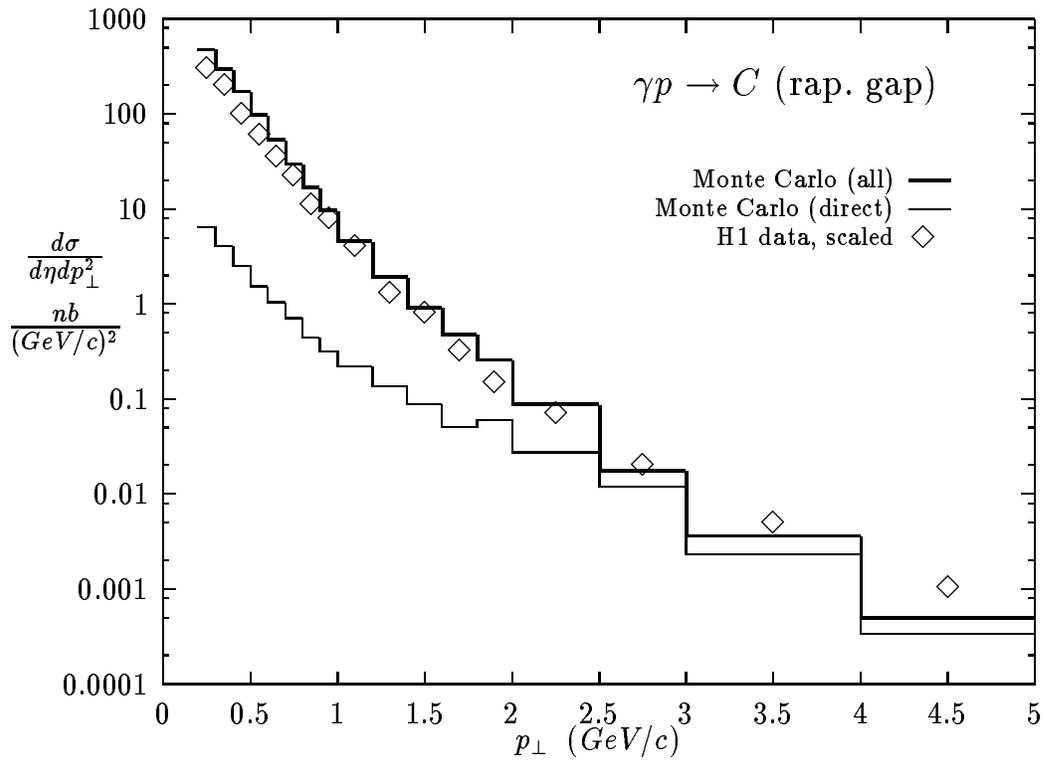
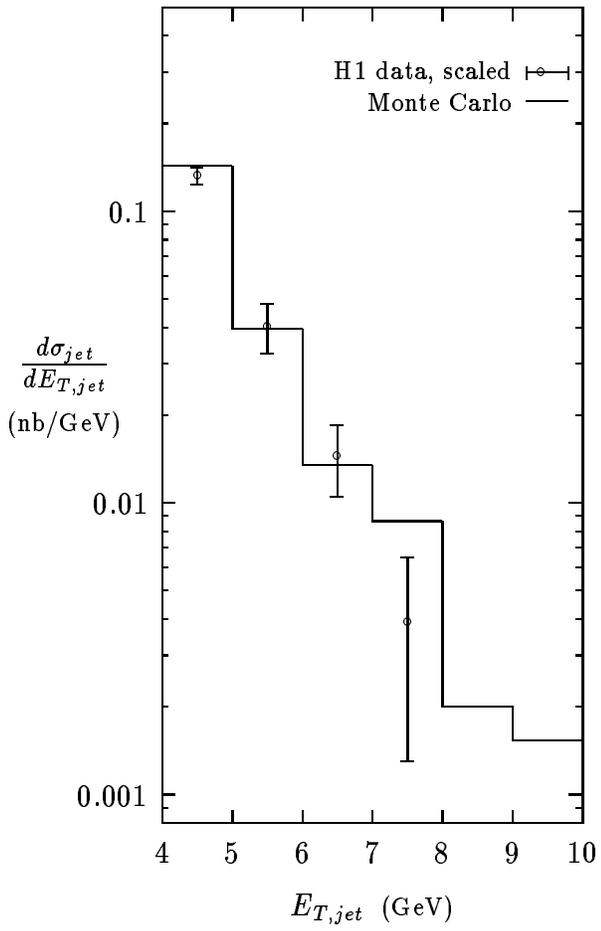
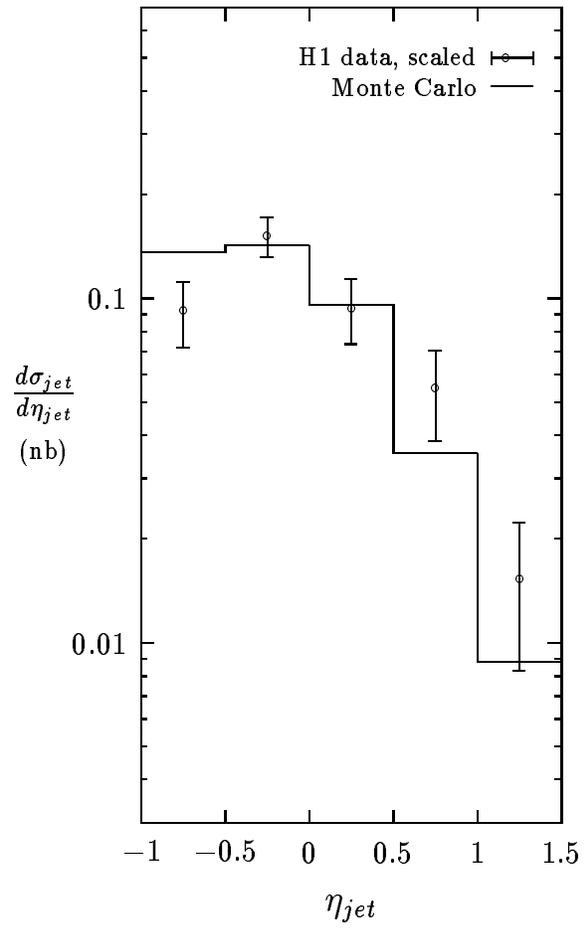
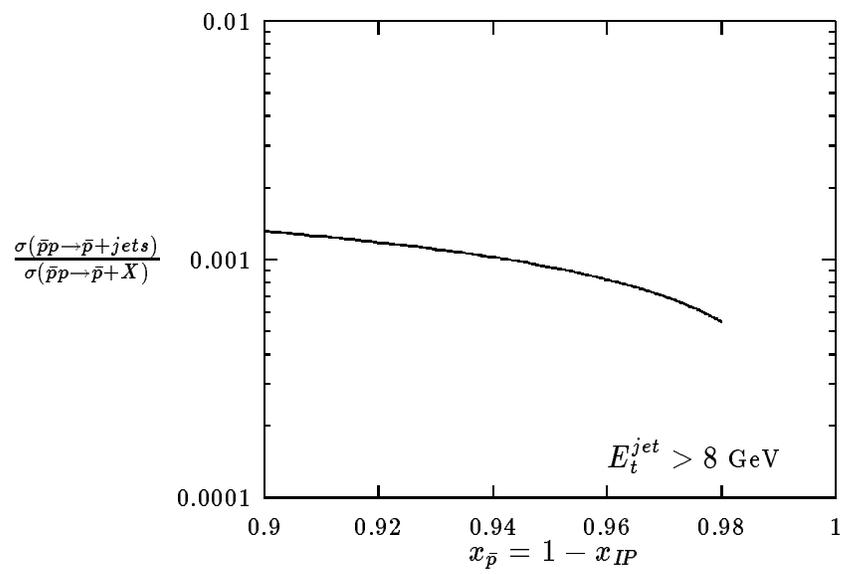
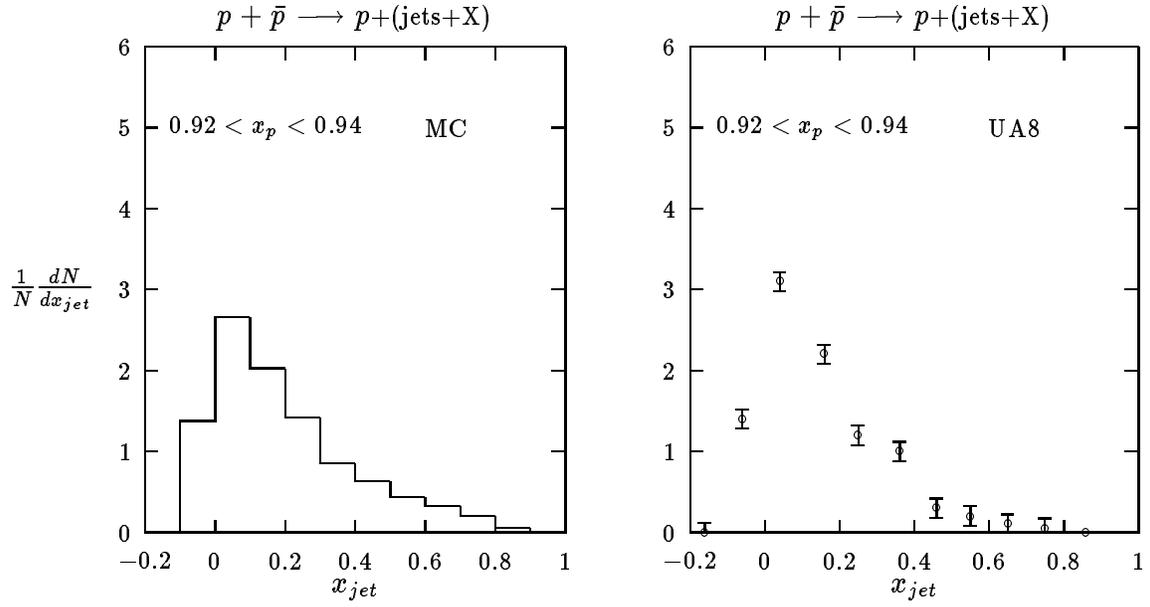
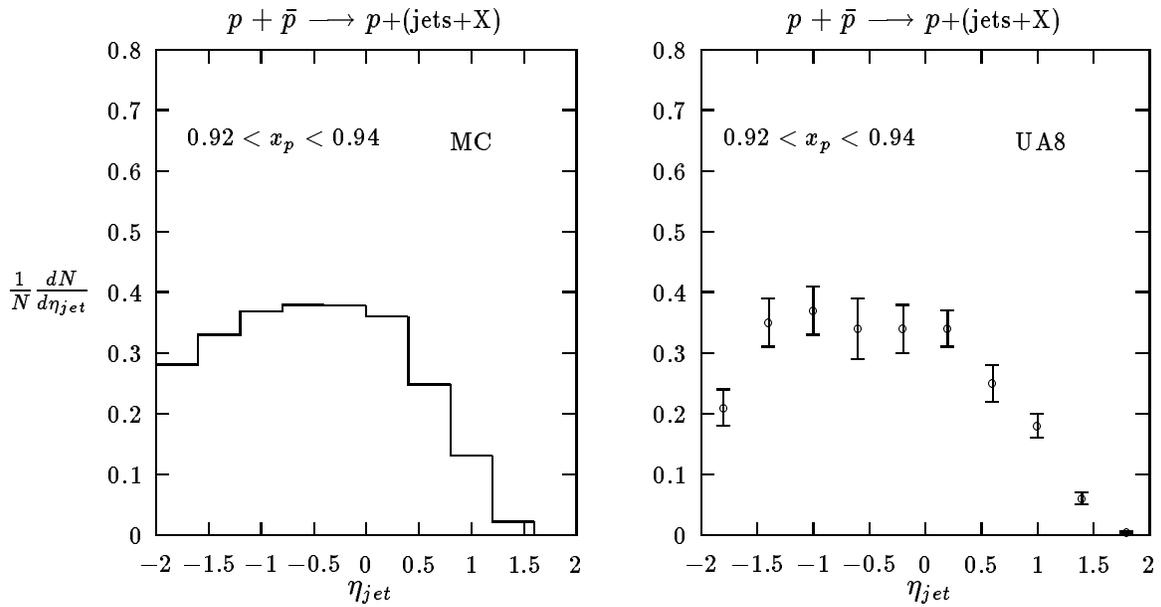
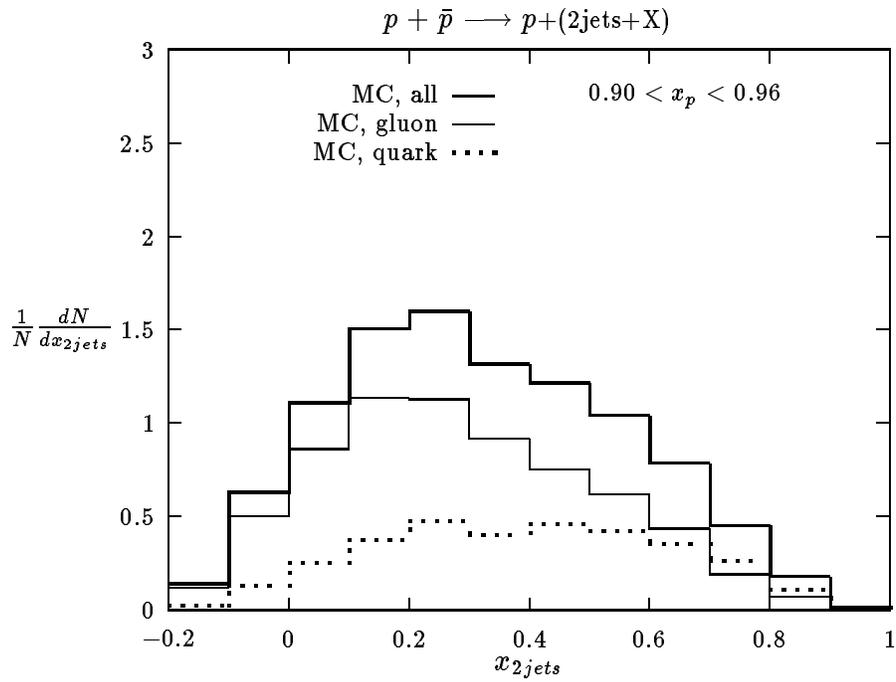
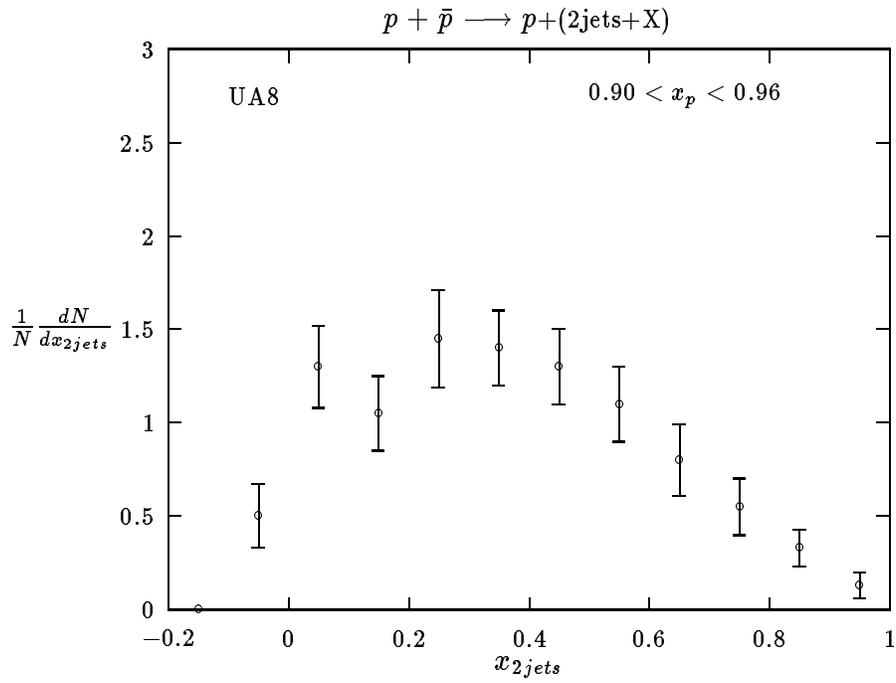
\end{document}